\documentclass[pra,twocolumn]{revtex4-1}
\usepackage{dcolumn}
\usepackage{bm}
\usepackage{amsmath,amssymb}
\usepackage{bm}
\usepackage{graphicx}
\usepackage{ascmac}
\usepackage[latin1]{inputenc}
\usepackage{braket}
\usepackage{txfonts}
\usepackage{mathrsfs}

\def\be{\begin{equation}}  
\def\ee{\end{equation}}     
\def\I{\mathrm{i}}  
\def\tr{\mathrm{tr}}
\def\e{\mathrm{e}}

\def\JS{J_\mathrm{S}}
\def\JR{J_\mathrm{R}}
\def\JQ{J_\mathrm{Q}}

\begin{document}

\title{Error trade-off relations for two-parameter unitary model with commuting generators}

\author{Shin Funada}
\author{Jun Suzuki}
\affiliation{
Graduate School of Informatics and Engineering, The University of Electro-Communications, 
1-5-1 Chofugaoka, Chofu-shi, Tokyo, 182-8585 Japan
}

\date{\today}

\begin{abstract}
We investigate whether a trade-off relation between the diagonal elements of the mean square error matrix exists for the two-parameter unitary models with mutually commuting generators. We show that the error trade-off relation which exists in our models of a finite dimension system is a generic phenomenon in the sense that it occurs with a finite volume in the spate space. We analyze a qutrit system to show that there can be an error trade-off relation given by the SLD and RLD Cramer-Rao bounds that intersect each other. First, we analyze an example of the reference state showing the non-trivial trade-off relation numerically, and find that its eigenvalues must be in a certain range to exhibit the trade-off relation. For another example, one-parameter family of reference states, we analytically show that the non-trivial relation always exists and that the range where the trade-off relation exists is up to about a half of the possible range.
\end{abstract}

\maketitle
\section{\label{sec:level1} Introduction \protect}
An error trade-off relation upon estimating quantum parametric models is significantly different from 
the case of classical statistics. There were many examples exhibiting such genuine quantum effects 
\cite{helstrom,holevo,yl73,RDGill,gibilisco,watanabe,suzuki:ijqi,RJDD16,kull}. 
The usual setting is to estimate the expectation values of two observables, 
thereby one can compare the derived error trade-off relation to the Heisenberg uncertainty relation.
Those previous studies on the trade-off relation focused on the settings where 
the observables are non-commuting.   

In the recent paper \cite{sf}, we gave an example of physical system that shows 
a non-trivial trade-off relation between estimation error for the expectation values of two commuting observables. 
We investigated the uncertainty relation, or trade-off relation obtained by estimating the position of an electron in a uniform magnetic field 
as a parameter estimation problem of two-parameter unitary model. 
In this model, shifts in the position of the electron was generated by a unitary transformation 
with the canonical momenta, $p_x$ and $p_y$.  
According to quantum mechanics, these generators $p_x$ and $p_y$ commute. 
  As the main conclusion of our paper \cite{sf}, 
we obtained a trade-off relation between $x$ and $y$, even though the generators of the unitary transformation, $p_x$ and $p_y$ commute. 

At first sight, this result came out counterintuitive, since two commuting generators should not 
give any correlation between two parameters of the quantum state. 
However, we clarified that two parameters are correlated, and hence, 
we cannot ignore a trade-off relation for our example. 
To be more precise, two symmetric logarithmic derivative (SLD) operators 
do not commute in this model, and hence it is genuine quantum mechanical.
A natural question is then whether error trade-off relations of this kind exist or not in general, 
in particular, models of finite dimensional systems. 
In Ref.~\cite{kull}, for example, the trade-off relation of qubit systems and qutrit systems were investigated. 
However, neither the existence of the intersection of the SLD and RLD bounds nor its contribution to the trade-off was discussed.

A key observation in our study is that the SLD Cram\'er-Rao (CR) inequality 
does not give any trade-off relations, whereas the right logarithmic derivative (RLD) CR inequality does. 
Importantly, these two CR bounds need to intersect each other in order to show a meaningful trade-off relation unless the RLD CR bound dominates the SLD CR bound. 
In this way, we can characterize the shape of the error trade-off relation more accurately.  

In this paper, we analyze finite dimensional systems to show that there can be such a trade-off relation given by 
the SLD and RLD CR bounds that intersect each other.  
As explicit examples, we study qutrit systems to demonstrate this counterintuitive result. 
We first disprove the existence of such error trade-off relations when the reference state is arbitrary pure states (Sec.~\ref{sec:purestate}) or mixed qubit states (Sec.~\ref{sec:purestate}). 
We then analyze error trade-off relations for a qutrit system numerically by randomly generating reference states. 
We observe that the occurrence of error trade-off relations is related to the eigenvalues of the reference state. 
In particular, they have to be distributed equally, otherwise a one particular large, or small, 
eigenvalue implies no error trade-off relation. 
To gain more insight into this kinds of trade-off relations, we show analytically that a non-trivial trade-off relation exists in 
a certain range of the reference state parameter which characterizes the reference state 
and that the region with the trade-off relation is up to about a half of the allowed region in one of those models.
\section{Model and error trade-off relation}
\subsection{Model}
Let us consider arbitrary finite dimensional system. 
We consider the two-parameter unitary transformation with the generators $X$ and $Y$ , i.e.,
\be
U(\theta^1, \, \theta^2)= \e^{-\I X \theta^1 - \I Y \theta^2}. \label{eq:unitary_def}
\ee
We denote the two-parameter family of states generated from the state $\rho_0$ as $\rho_\theta$. 
\be
\rho_\theta = U(\theta^1, \, \theta^2) \, \rho_0 U^\dagger (\theta^1, \, \theta^2). \label{eq:transform}
\ee
The state $\rho_0$ is called as a reference state. 
In this paper, we mainly consider the case of the commuting generators, $[X, \, Y]=0$ unless stated explicitly.
\subsection{Error trade-off relation}
In the remaining of the paper, we consider unitary models only. 
We shall drop the parameter $\theta$ to denote the quantum Fisher information matrices, 
since they are independent of the parameter. 
To derive a trade-off relation between the diagonal components of the mean square error (MSE) matrix $V=[V_{ij}]$, 
suppose we have a quantum CR inequality 
\[
V\ge \JQ^{\: -1},
\]
with $\JQ$ a quantum Fisher information matrix. 
In Ref.~\cite{sf}, we derived a trade-off relation based on the inequality below,
\be
\left[V_{11} - \JQ^{\: 11} \right]\left[V_{22} - \JQ^{\: 22} \right] > \left| \mathrm{Im} \, \JQ^{\: 12} \right|^{\,2}. \label{tradeoff}
\ee
where $\JQ^{-1}=[\JQ^{\: ij}]$. $ \mathrm{Im}$ denotes the imaginary part of a complex number. 
From this expression, we see that there exists a non-trivial error trade-off relation when $\mathrm{Im} \, \JQ^{\: 12} \neq 0$. 
Note, however, that this inequality alone does not give a conclusive argument whether an error trade-off relation exists or not. 
This is because the quantum CR inequality is not tight unless certain special conditions are satisfied. 
The central idea of this paper is to consider two different CR inequalities set by the SLD and RLD Fisher information matrices. 
When combining two error trade-off relations, we can determine the shape of an error trade-off relation more accurately. 

From the discussion above, we do not have the trade-off relation given by Eq.~\eqref{tradeoff} when 
the SLD Fisher information matrix $\JS$ is used. 
This is because it is a real symmetric matrix. 
The other candidate for giving rise to an error trade-off relation is the RLD Fisher information matrix $\JR$. 
In this case, the necessary condition to have an error trade-off relation is 
\be
\left| \mathrm{Im} \, \JR^{\:12} \right|^{\,2} \neq 0. \label{tradeoff2}
\ee
as in Eq.~\eqref{tradeoff}.  
By defining
$\delta:=J_{\mathrm{R}, \, 12} - J_{\mathrm{R}, \, 21}$, we have an equivalent condition, 
\be \label{cond:Im}
\mathrm{Condition \, 1}: \qquad \delta=J_{\mathrm{R}, \, 12} - J_{\mathrm{R}, \, 21} \neq 0,
\ee
where $\JR=[J_{\mathrm{R}, \, ij}]$. 

$(i, \, j)$ component of the RLD Fisher information matrix, 
$J_{\mathrm{R}, \, ij}$ is defined as follows. 
\be
J_{\mathrm{R}, \, ij}= \tr\left(\rho_0 L_{\mathrm{R}, \, j} \, L^\dagger_{\mathrm{R}, \, i}  \right),
\ee
where $\partial_i \rho_\theta \, |_{\theta=0} = \rho_0 L_{\mathrm{R}, \, i}$. 
By using $\partial_i \rho_\theta \, |_{\theta=0} = L^\dagger_{\mathrm{R}, \, i} \, \rho_0$, Eqs~\eqref{eq:unitary_def}, and ~\eqref{eq:transform}, we obtain 
\be
J_{\mathrm{R}, \, ij} = - \tr\left([X_j \, \rho_0] [X_i, \, \rho_0] \, \rho_0^{-1}\right), \label{RLD}
\ee
where $X_1=X$ and $X_2=Y$. 
With this, Condition 1 is 
\begin{equation} 
\delta = \tr\left(\big[ [X, \, \rho_0]\,,\, [Y, \, \rho_0]\big] \, \rho_0^{-1}\right) , \label{eq:delta}
\end{equation}
and thus, it is relatively easy to check this condition analytically. 
We stress that having commuting generators, $[X,Y]=0$ does not immediately imply $\delta=0$. 

Now suppose Condition 1 is satisfied. 
\begin{equation}
\left[V_{11} - J_\mathrm{R}^{\:11}\right]\left[V_{22} - J_\mathrm{R}^{\:22}\right] > \left| \mathrm{Im} \, J_\mathrm{R}^{\:12} \right|^{\,2}. \label{tradeoff_RLD}
\end{equation}
To give a conclusive argument for the existence of an error trade-off relation, 
we also consider consider the SLD CR inequality. 
Since the SLD Fisher information matrix is real, 
the diagonal components of the MSE matrix obey
\begin{equation}
V_{11} - J_\mathrm{S}^{\:11}\ge0,\ V_{22} - J_\mathrm{S}^{\:22}\ge0. \label{tradeoff_SLD}
\end{equation}
Note that from the general relationship between the SLD and RLD Fisher information matrices, we have~\cite{petz}
\be
\JS^{\:-1} \ge \mathrm{Re} \, (\JR^{\: -1}), \nonumber
\ee
where $\mathrm{Re}$ denotes the real part of a matrix. 
Since there exists a locally unbiased estimator such that its MSE for $\theta^1$ 
is arbitrary close to $J_\mathrm{S}^{\:11}$, (The same statement holds for $\theta^2$ as well.)
intersections of two inequalities \eqref{tradeoff_RLD} and \eqref{tradeoff_SLD} 
imply the existence of the error trade-off relation. 
See Fig.~\ref{fig5} for the occurrence of intersections of the two bounds 
and the error trade-off relation as an example.  
Working out elementary algebra, 
we find that the following condition needs to be satisfied 
in order to have a non-trivial error trade-off relation~\cite{sf}. 
\be\label{cond:intersec}
\mathrm{Condition \, 2}: \quad \Delta := \left|\mathrm{Im} \, J_\mathrm{R}^{\:12} \right|^2 - \left[J_\mathrm{R}^{\:11} - J_\mathrm{S}^{\:11}\right]\left[J_\mathrm{R}^{\:22} - J_\mathrm{S}^{\:22}\right] > 0. 
\ee 
And hence, we have a trade-off relation between $V_{11}$ and $V_{22}$ if these two conditions 
\eqref{cond:Im} and \eqref{cond:intersec} are satisfied. 

Next, let us make a remark about D-invariant models. 
It is known that, the RLD CR inequality is saturated when the model is D-invariant 
\cite{kahn,hayashi,kahn2,yamagata,suzuki,yang}. This is true at least in the asymptotic setting. 
There is no intersection of the RLD and SLD CR bounds in the D-invariant models, 
because the RLD CR bound is dominant over the SLD CR bound.  
If the model is D-invariant and if the imaginary part of the off-diagonal elements of RLD Fisher information matrix are not zero, there is a trade-off relation that results from Condition 1 only. 
In the following, we mainly investigate the non-asymptotic setting unless stated explicitly. 
This is in contrast to the previous study~\cite{kull}, where the authors focused on the D-invariant model. 

\section{Reference state: Pure state} \label{sec:purestate}
We first consider the case that the reference state is a pure state, i.e., $\rho_0=\ket{\psi_0} \bra{\psi_0}$. 
From Eqs.~\eqref{eq:unitary_def} and~\eqref{eq:transform},  
$\rho_\theta$ is expressed as
\be
\rho_\theta= \e^{-\I X \theta^1 - \I Y \theta^2} \ket{\psi_0} \bra{\psi_0}  \e^{\I X \theta^1 + \I Y \theta^2}.
\ee
Therefore, we have
\begin{align}
\partial_1\ket{\psi_\theta} =  -\I X \ket{ \psi_\theta}, \\
\partial_2\ket{\psi_\theta} =  -\I Y \ket{ \psi_\theta},  
\end{align}
where $\ket{\psi_\theta} = \e^{-\I X \theta^1 - \I Y \theta^2} \ket{\psi_0}$. 
In the pure state model, the RLD does not exist in general. 
Here, we use the generalized RLD instead \cite{fujiwara2}.
The components of generalized RLD Fisher information matrix, $\tilde{J}_{\mathrm{R}, \,12}$ and 
$\tilde{J}_{\mathrm{R}, \,21}$ are given by \cite{fujiwara}
\begin{align}
\tilde{J}_{\mathrm{R}, \,12}
&=4(\braket{ \psi_0 | Y\, X | \psi_0} - \braket{\psi_0 | Y | \psi_0} \braket{\psi_0 | X | \psi_0}), \\
\tilde{J}_{\mathrm{R}, \, 21}
&=4(\braket{ \psi_0 | X\, Y | \psi_0} - \braket{\psi_0 | X | \psi_0} \braket{\psi_0 | Y | \psi_0}).
\end{align}
We obtain $\delta$ in Condition 1, or Eq.~\ref{cond:Im} as follows.
\be
\delta =\tilde{J}_{\mathrm{R}, \,12} - \tilde{J}_{\mathrm{R}, \,21} = 4 \braket{ \psi_0 | [Y, \, X] | \psi_0}.
\ee
If the reference state is a pure state and if the generators commute, Condition 1 does not hold. 
Therefore, there is no trade-off relation given by Eq.~\eqref{tradeoff}. 
See also ~\cite{matsumoto}. 
\section{Reference state: Qubit state}
\subsection{General case}
We consider the case of a single qubit in a mixed state. 
We first consider the general two-parameter unitary model to get insight into the problem. 
By using the Bloch vector, we can express the reference state $\rho_0$ as
\be
\rho_0 = \frac{1}{2}(\mathrm{I} + \vec{s}_0 \cdot \vec{\sigma}).
\ee
where $|\vec{s}_0|<1$.  $\rho_\theta$ is given by Eq.~\eqref{eq:transform}.
The generators $X, \, Y$ can also be expanded with using Pauli matrices.
\begin{align}
X &= x_{0} \, \mathrm{I} + \vec{x} \cdot \vec{\sigma},  \label{eq:pauli_x} \\
Y &= y_{0} \, \mathrm{I} + \vec{y} \cdot \vec{\sigma}. \label{eq:pauli_y} 
\end{align}
The inverse of SLD and RLD Fisher information matrices, $\JS^{\:-1}$ and $\JR^{\:-1}$ are explicitly written as 
\begin{align}
\JS^{\:-1}&= \frac{4}{\mathrm{det} \, \JS} \begin{pmatrix}
 ( \vec{y} \times  \vec{s}_0)^2 &  -( \vec{x} \times  \vec{s}_0) \cdot  ( \vec{y} \times  \vec{s}_0)\\
 - ( \vec{x} \times  \vec{s}_0) \cdot  ( \vec{y} \times  \vec{s}_0) &   ( \vec{x} \times  \vec{s}_0)^2 \\
\end{pmatrix}, \label{eq:qubit_SLDinv}\\
\JR^{\:-1}&=  \JS^{-1}  \nonumber \\
&+\frac{4}{\mathrm{det} \, \JS}  \begin{pmatrix}
 0 & -\I  \vec{s}_0^{\:2} [ \vec{s}_0 \cdot (\vec{x} \times \vec{y} ) ]\\
  \I \vec{s}_0^{\:2} [ \vec{s}_0 \cdot (\vec{x} \times \vec{y} ) ] &  0 
\end{pmatrix}, \label{eq:qubit_RLDinv}
\end{align}
where $\mathrm{det} \, \JS$ is the determinant of $\JS$, and it is
\be
\mathrm{det} \, \JS={16\, \vec{s}_0^{\:2} [ \vec{s}_0 \cdot (\vec{x} \times \vec{y} ) ]^2} .
\ee
As shown in Eqs.~\eqref{eq:qubit_SLDinv} and~\eqref{eq:qubit_RLDinv}, $\JS^{\:-1}=\mathrm{Re} \, \JR^{\:-1}$ holds. 
It follows that our qubit model is D-invariant.  (See Lemma III-3 in Ref.\cite{suzuki}.)
Therefore, the RLD CR bound is asymptotically achievable and gives a trade-off relation. 
In this case, as explained earlier, the SLD and RLD CR bounds do not have intersections, but the trade-off relation exists in the asymptotic setting. 

As for the Nagaoka bound, or the Gill-Massar (GM) bound for a two-parameter qubit model, 
which is known to be achievable \cite{nagaoka2,RDGill} in the non-asymptotic setting, 
an inequality regarding the diagonal components of the MSE matrix can be
derived.  The inequality of the Nagaoka band is written as
\be
\left[V_{11} - \JS^{\: 11} \right]\left[V_{22} - \JS^{\: 22} \right] > \frac{1}{\mathrm{det} \, \JS}.
\ee
From Eqs.~\eqref{tradeoff_SLD} and~\eqref{eq:qubit_RLDinv}, we obtain the inequality of the RLD CR bound. 
\be
\left[V_{11} - \JS^{\: 11} \right]\left[V_{22} - \JS^{\: 22} \right] > \frac{\vec{s}_0^{\:2}}{\mathrm{det} \, \JS}.
\ee
We used $\JR^{\:11}=\JS^{\:11}$ and $\JR^{\:22}=\JS^{\:22}$. Since $\vec{s}_0^{\:2} < 1$, 
the Nagaoka bound is tighter than the RLD CR bound. 
This is because the Nagaoka bound is achieved by a separable measurement. 

\subsection{Commuting generators' case}
Next, we derive a relationship between $X$ and $Y$, or $\vec{x}$ and $\vec{y}$ when  $X$ and $Y$ commute. 
From Eqs.~\eqref{eq:pauli_x} and~\eqref{eq:pauli_y}, the commuting relation of $X$ and $Y$ is given as
\be
[X, \, Y] = [\vec{x} \cdot \vec{\sigma}, \, \vec{y} \cdot \vec{\sigma}] = 2 \I (\vec{x} \times \vec{y})\cdot \vec{\sigma}. \nonumber 
\ee
It immediately follows that the necessary and sufficient condition for $X$ and $Y$ to commute is 
$\vec{x} \times \vec{y} = \vec{0}$. 
There is no trade-off relation because the unitary transformation is no longer two-parameter model, 
because $\vec{x}$ and $\vec{y}$ are parallel when $X$ and $Y$ commute. 
\section{Reference state: Qutrit state}
Let us consider a qutrit system, the three-dimensional system. 
To avoid non-regular models, we consider the full-rank model. 
Other regularity conditions are also imposed implicitly. 

Since $X$ and $Y$ commute, they are simultaneously diagonalizable.  
Without the loss of generality, for the calculation of $\delta$, 
we can use the representation so that both $X$ and $Y$ can be diagonalized .
\begin{align}
\rho_0 &=\begin{pmatrix}
\rho_{11} & \rho_{12} & \rho_{13} \\
\rho_{21} & \rho_{22} & \rho_{23} \\
\rho_{31} & \rho_{32} & \rho_{33} \\
\end{pmatrix}, \\
X&=\begin{pmatrix}
x_{1} & 0 &0 \\
0& x_{2} & 0 \\
0 & 0 & x_{3} \\
\end{pmatrix}, \nonumber \\
Y&=\begin{pmatrix}
y_{1} & 0 &0 \\
0& y_{2} & 0 \\
0 & 0 & y_{3} \\
\end{pmatrix} . 
\end{align}
By using Eq.~\eqref{RLD}, $\delta$ is calculated as follows.
\begin{align}
\delta
&=(\det\rho_0)^{-1}\, \big(\rho_{12}\rho_{23}\rho_{31} - \rho_{21}\rho_{32}\rho_{13}\big)\,\left[\big(\vec{y} \times \vec{x}\big)\cdot \vec{1}\right],
\end{align}
where $\vec{x}=(x_{1}, x_{2}, x_{3} ), \: \vec{y}=(y_{1}, y_{2}, y_{3}  ),$ and $\vec{1}=(1, 1, 1 )$.
The condition of no trade-off relation, $\delta=0$ holds when
\begin{align}
\mathrm{Im}\left( \rho_{12}\rho_{23}\rho_{31} \right) &= 0, \label{rho} \\
 \mathrm{or} \nonumber\\
(\vec{y} \times \vec{x})\cdot \vec{1} &= 0 . \label{eq:xy}
 \end{align}
Violation of these conditions together with Eq.~\eqref{cond:intersec} are the necessary and sufficient 
conditions to have a non-trivial error trade-off relation. In the case of qutrit, we cannot give an explicit expression 
of $\Delta$ in general.  
But, we can obtain $\Delta$ in a straightforward manner numerically. 

In the following subsections, we give examples of reference states that give non-trivial error trade-off relations. 
One of them gives a relatively high possibility. 
Our main interest is to investigate the error trade-off relation for a given commuting $X$ and $Y$. 
\subsection{Example: reference state with multi-parameter } \label{sec:q_model}
As one of the simplest examples, we pick an example with pure imaginary off-diagonal 
components as a reference state $\rho_0$ with five reference state parameters 
$v_1$,  $v_2$, $v_3$, $u_1$,  $u_2$, and  $u_3$. ($v_1+v_2+v_3=1$) 
\be
\rho_0=\frac{1}{3}
\begin{pmatrix}
 v_1 & - \I \sqrt{u_1} & \I \sqrt{u_2} \\
  \I \sqrt{u_1} &    v_2 & - \I \sqrt{u_3} \\
 - \I \sqrt{u_2} & \I \sqrt{u_3} &    v_3 \\
\end{pmatrix}.  \label{ref_state}
\ee
We choose the reference state $\rho_0$ as above, because imaginary parts of the off-diagonal components of the reference state $\rho_0$ 
are important to satisfy Condition 1 as seen in Eq.~\eqref{rho}. 
\begin{figure}[t]
\begin{center}
\includegraphics[width=6cm]{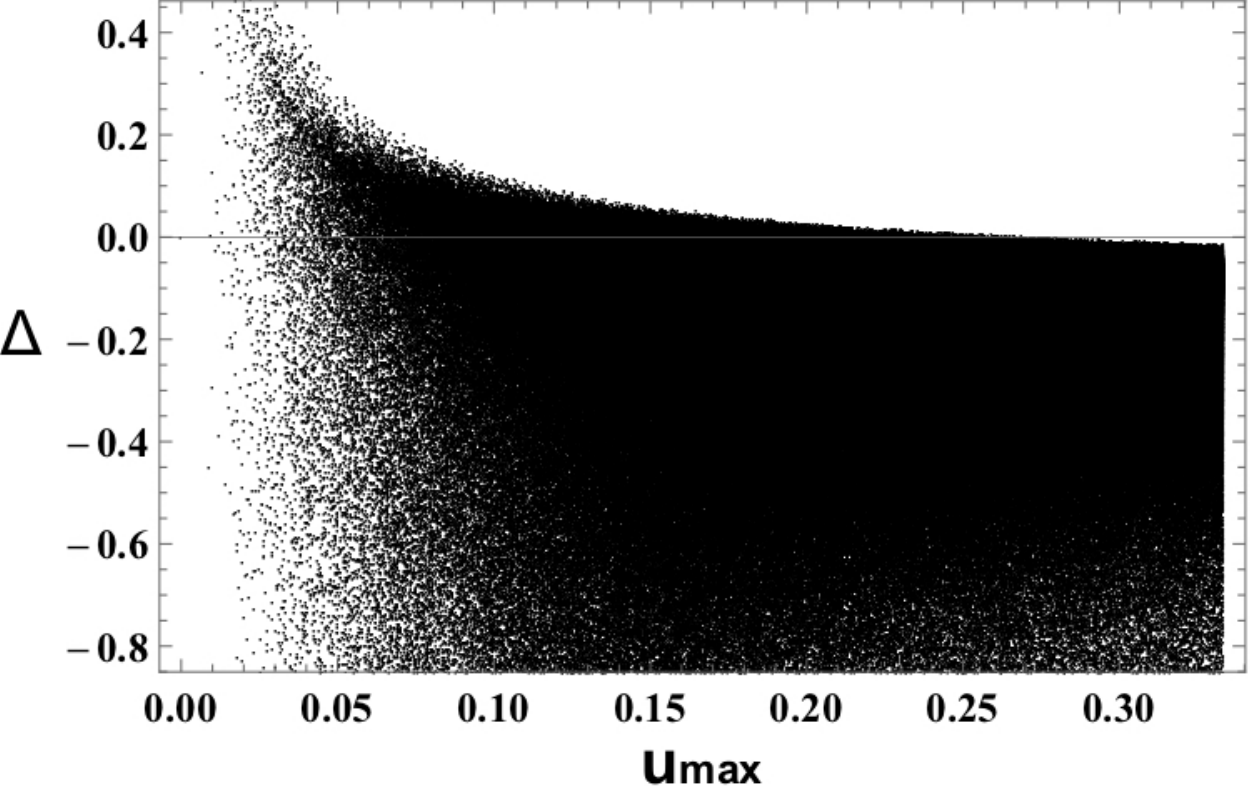}
\caption{$\Delta$ as a function of $u_{max}$, the maximum of $u_1, \, u_2$, and $u_3$ in Eq.~\eqref{ref_state}. 
$\vec{x}=(1,~2,~3 )$, $\vec{y}=(1.5,~5,~1)$.}
\label{fig1}
\end{center}
\end{figure}
We calculate $\Delta$ in Condition 2 with using the reference state $\rho_0$ defined by Eq.~\eqref{ref_state} 
of which reference state parameters are generated by random numbers. We pick those which satisfy $\tr \, \rho_0=1$ and $\rho_0 > 0$ 
and calculate the RLD and SLD Fisher information matrices $\JS$ and $\JR$. The RLD Fisher information matrix is obtained by using 
Eq.~\eqref{RLD}. The SLD Fisher information calculation is done in the standard method. 
(See for example, Refs.~\cite{paris,Liu}.) 
The number of samples generated is on the order of $10^6$. 
Figure~\ref{fig1} shows $\Delta$ as a function of $u_{max}$, the maximum of $u_1$, $u_2$, and $u_3$.  There exists a region $\Delta>0$. 
The ratio of obtaining $\Delta>0$ out of all of the samples generated is 3.0\%.
Figures~\ref{fig2} and~\ref{fig3} show $\Delta$ as a function of $\lambda_{min}$ and $\lambda_{max}$, respectively.  
$\lambda_{min}$ and $\lambda_{max}$ are the minimum and maximum of eigenvalues of $\rho_0$, respectively.  For $\Delta$ to be positive, 
$\lambda_{min}$ and $\lambda_{max}$ must be in a certain range. $\lambda_{min}$ is more than about 0.13 and 
$\lambda_{max}$ is less than about 0.58.
\begin{figure}[t]
\begin{center}
\includegraphics[width=6cm]{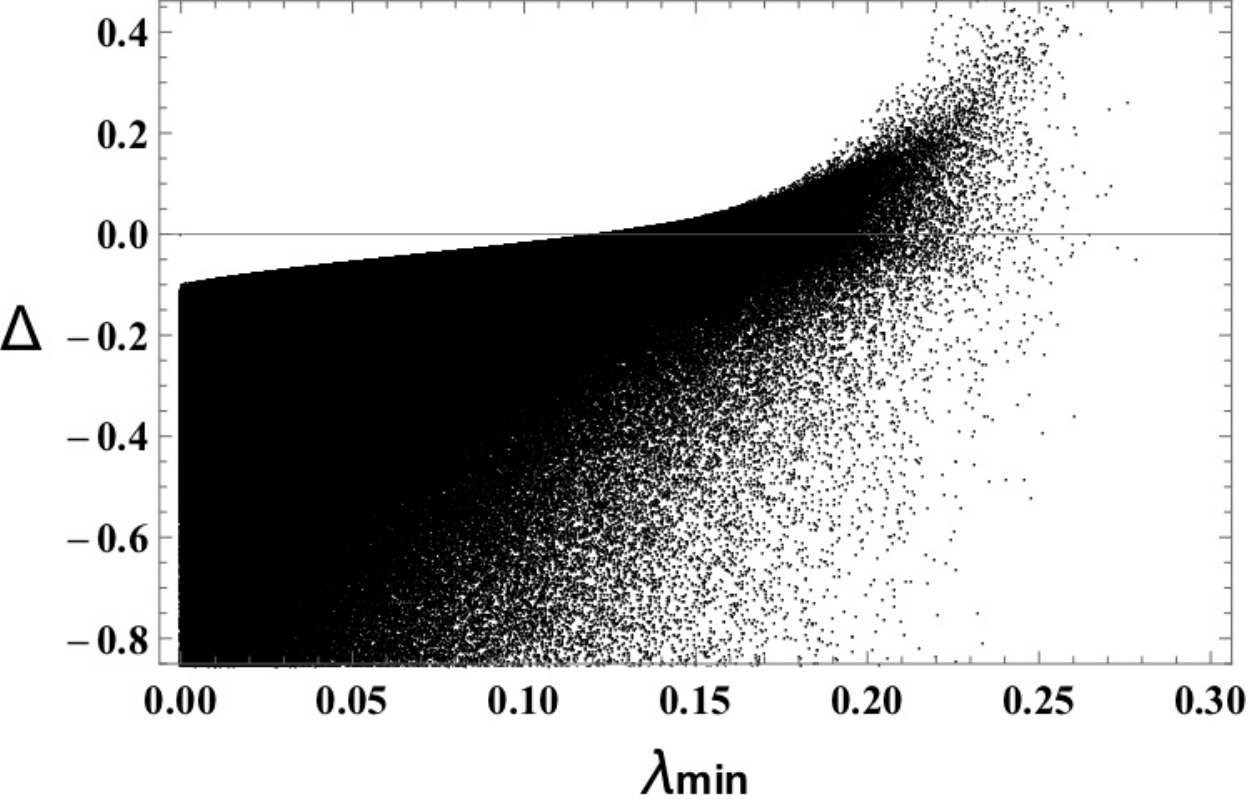}
\caption{$\Delta$ as a function of $\lambda_{min}$, the minimum of the eigenvalues of $\rho_0$}
\label{fig2}
\end{center}
\end{figure}
\begin{figure}[t]
\begin{center}
\includegraphics[width=6cm]{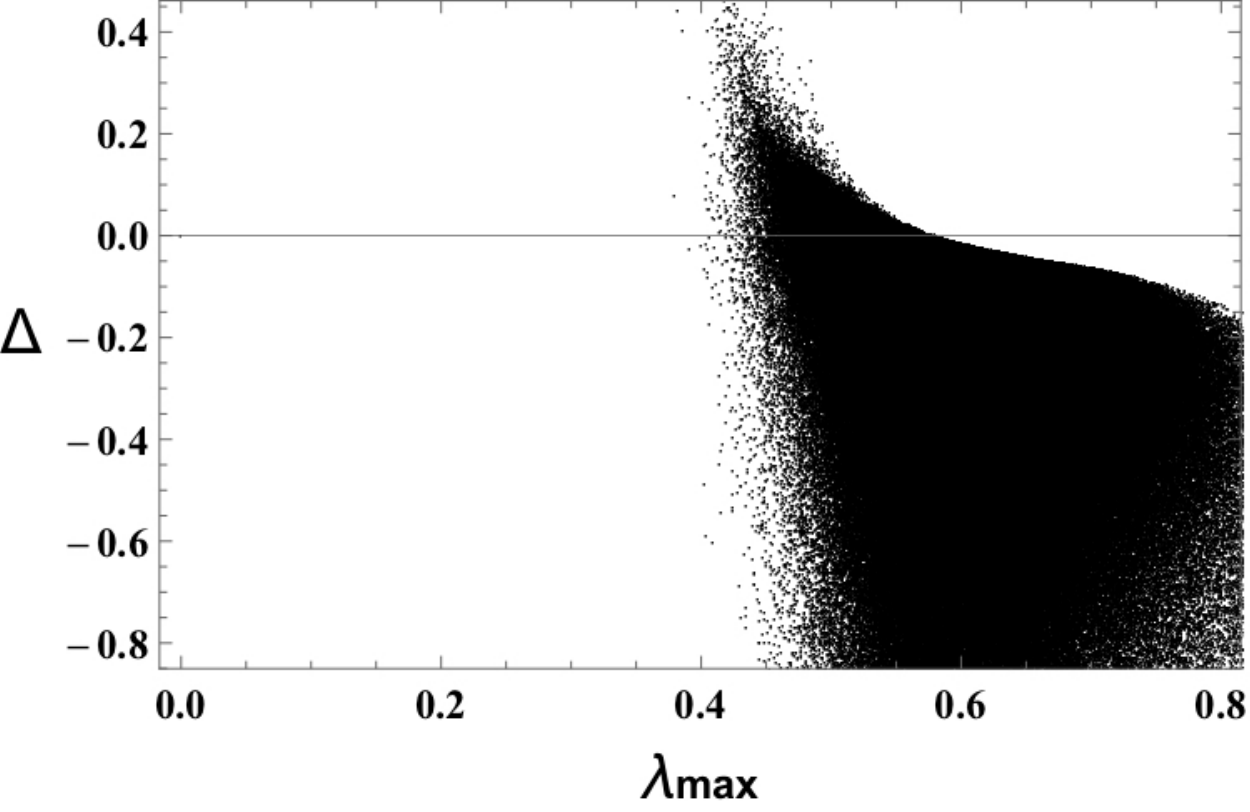}
\caption{$\Delta$ as a function of $\lambda_{max}$, the maximum of the eigenvalues of $\rho_0$}
\label{fig3}
\end{center}
\end{figure}
\subsection{Example: 
one-parameter family of reference states} \label{one_u} 

Next, we set the reference state parameters in Eq.~\eqref{ref_state} as $v_1=v_2=v_3=1$ and $u_1=u_2=u_3=u$ 
in order to investigate the model more in detail analytically. 
We pick the reference state parameters as above, because the result of Section~\ref{sec:q_model} indicates that 
the eigenvalues of the reference state Eq.~\eqref{ref_state} be roughly in the range $1/3 \pm 0.2$ to exhibit the non-trivial trade-off relation. 
The reference state $\rho_0$ is, then explicitly written as
\be
\rho_0=\frac{1}{3} I + \frac{1}{3} \sqrt{u}
\begin{pmatrix}
 0 & - \I  & \I  \\
  \I &    0 & - \I \\
 - \I & \I  &  0 \\
\end{pmatrix}, \label{single_u}
\ee
where $I$ denotes 3$\times$3 identity matrix. The reference state $\rho_0$ is a sum of the completely mixed state of the qutrit system 
and a perturbation with one parameter $u$.  
The parameter $u$ must be in the range, $0<u<1/3$ for the reference state $\rho_0$ to be positive.
We exclude $u=0$, because $\rho_0= \mathrm{I}/3$ at $u=0$. 

In the following, we show that the reference state $\rho_0$ Eq.~\eqref{ref_state}
always gives a non-trivial trade-off relation with a certain choice of the reference state parameter $u$ and 
that the possibility of seeing the non-trivial trade-off relation is not small. 
\subsubsection{Intersections of RLD and SLD CR bounds}
From Condition 2 expressed by Eq.~\eqref{cond:intersec}, $\Delta > 0$ needs to be satisfied 
in order to have a non-trivial error trade-off relation.
We define a geometrical parameter, $\zeta$ as follows.
\be
\zeta=\frac{[ \vec{1} \cdot(\vec{x}\times \vec{y}) ] ^2}{(\vec{1}\times \vec{x})^2 (\vec{1}\times \vec{y})^2}. \label{t}
\ee
Let $\vec{\xi}=\vec{1} \times \vec{x}$ and $\vec{\eta}=\vec{1} \times \vec{y}$.  
A vector analysis formula gives an expression, 
\be
\zeta = \frac{1}{3} \sin^2 \theta \leq \frac{1}{3}, 
\ee
where $ \sin \theta = | \vec{\xi} \times \vec{\eta} | / ( |\vec{\xi} | |\vec{\eta}|)$.
$\zeta = 1/3$ when $\theta = \pm{\pi /2}$. $\zeta=0$ is excluded, 
because $\vec{\xi} \times \vec{\eta}=\vec{0}$ gives $\delta=0$ from Eq.~\eqref{eq:xy}. 
Therefore, the possible range for the parameter $\zeta$ is $0 < \zeta \leq 1/3$.

We introduce  
a function of $u$ at a given $\zeta$, $F_\zeta(u)$ as
 \be
 F_\zeta(u)= 16 \zeta (3 u^2-7 u+2)^2-u (3 u^2-9 u+8)^2. \label{gts}
\ee
By using $F_\zeta(u)$, $\Delta$ is expressed as
\be
\Delta= \frac{9}{16 \zeta^2 |\vec{\xi}|^2 |\vec{\eta}|^2 (2-u)u (u^2-7 u +4)^2} F_\zeta(u).
\label{Delta_gts}
\ee
The coefficient of $F_\zeta(u)$ in Eq.~\eqref{Delta_gts} is positive finite when $0 < u < 1/3$. 
In order to investigate the range of $u$ that gives $\Delta > 0$, we can check the condition for $F_\zeta(u) > 0$ instead.
 
We can analytically show that $F_\zeta(u)$ is a monotonically decreasing function of $u$ 
and that there is always a unique solution $u_0$ that satisfies $F_\zeta(u_0)=0$ 
when $0< \zeta \leq 1/3$ and when $\rho_0 > 0$, i.e., $0 <u<1/3$.  
A detailed explanation is given in Appendix~\ref{sec:sol_gts}. 
Figure~\ref{fig4} shows the solution $u_0$ that satisfies $F_\zeta(u)=0$. In the region where 
$u < u_0$ at a given $\zeta$, the non-trivial trade-off relation exists.  We can regard $u_0$ as 
the upper limit of $u$ that gives a non-trade off relation. 
It is worth noting that the upper limit of $u$ is almost a half of 
the maximum of $u$, $1/3$ at $\zeta = 1/3$.  This means that the possibility of realizing non-trivial trade-off relation is not small. 
\begin{figure}[t]
\begin{center}
\includegraphics[width=6.5cm,clip]{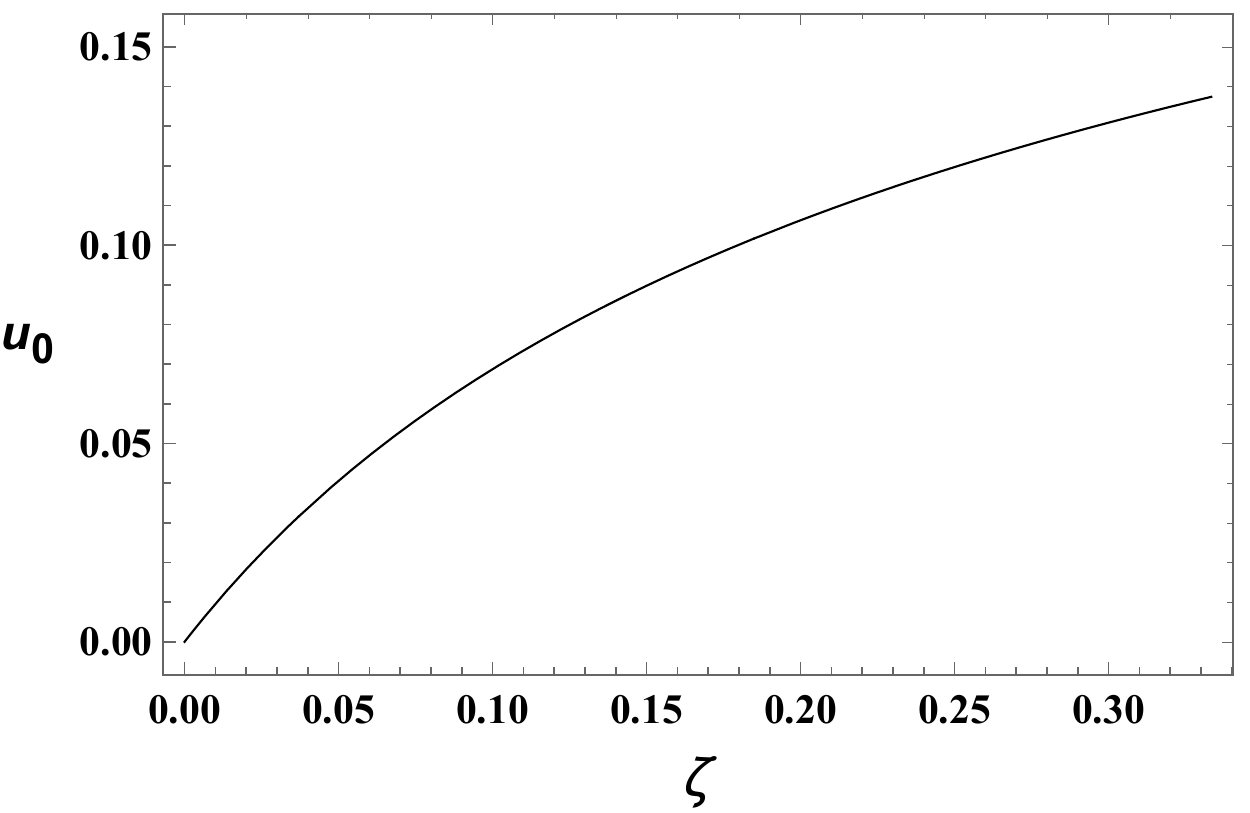}
\caption{Solutoin $u_0$ that satisfies $F_\zeta(u_0)=0$ in the range $0<\zeta \leq 1/3$}
\label{fig4}
\end{center}
\end{figure}

\begin{figure}[t]
\begin{center}
\vspace{0mm}
\includegraphics[width=7cm,clip]{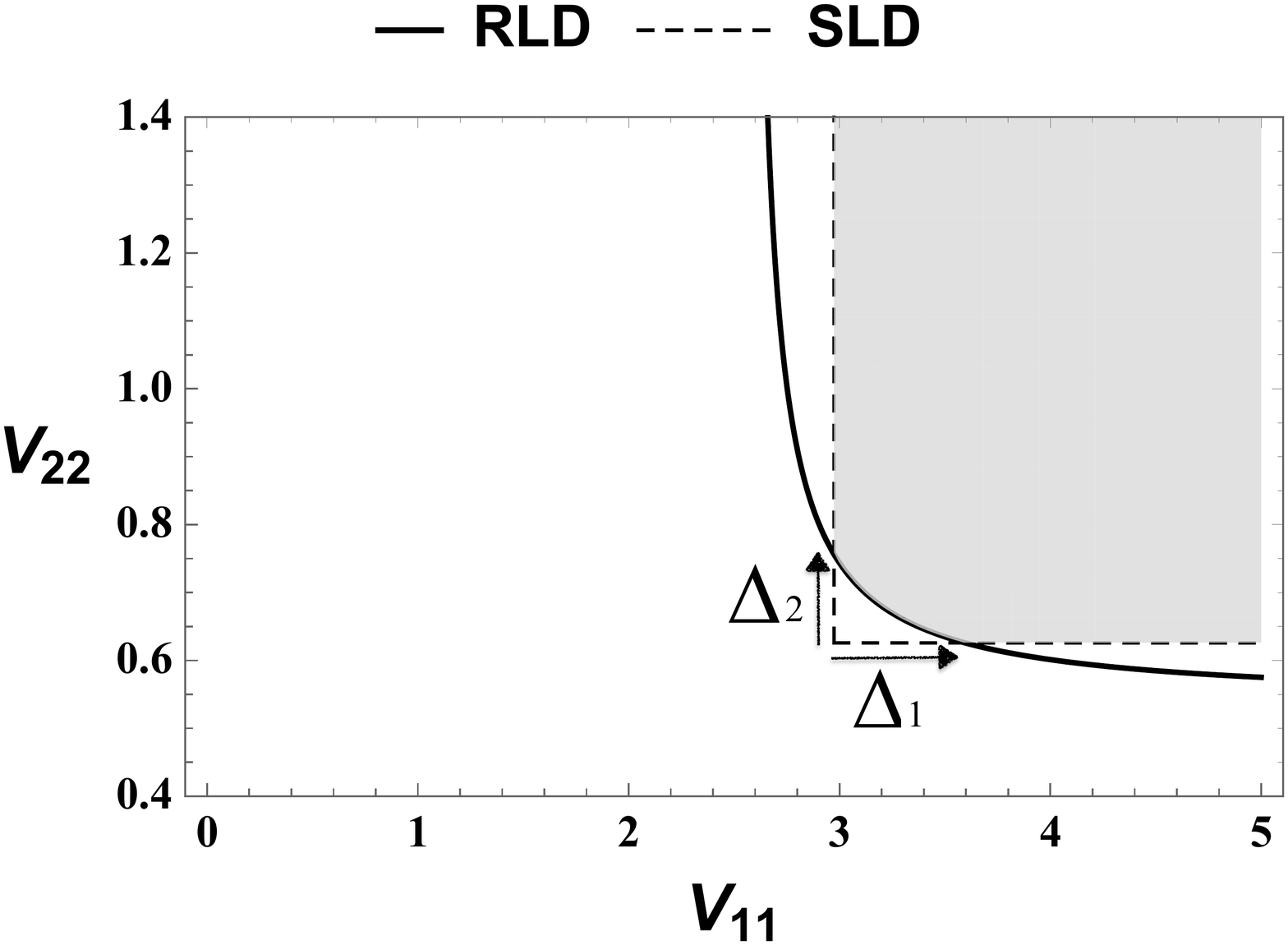}
\vspace{0mm}
\caption{Example of RLD and SLD CR bounds with the intersections: 
the reference state $\rho_0$ defined by Eq.~\eqref{single_u} with $u=1/12$, $\vec{x}=(1,~2,~3 )$, $\vec{y}=(1.5,~5,~1)$. 
The gray region is an allowed region.}
\label{fig5}
\end{center}
\end{figure}

Figure~\ref{fig5} shows an example in which the SLD and RLD CR bounds have two intersections.  
The parameters used are $u= 1/12$, $\vec{x}=(1,~2,~3 )$, and $\vec{y}=(1.5,~5,~1)$.  

\begin{figure}[t]
\begin{center}
\includegraphics[width=6cm]{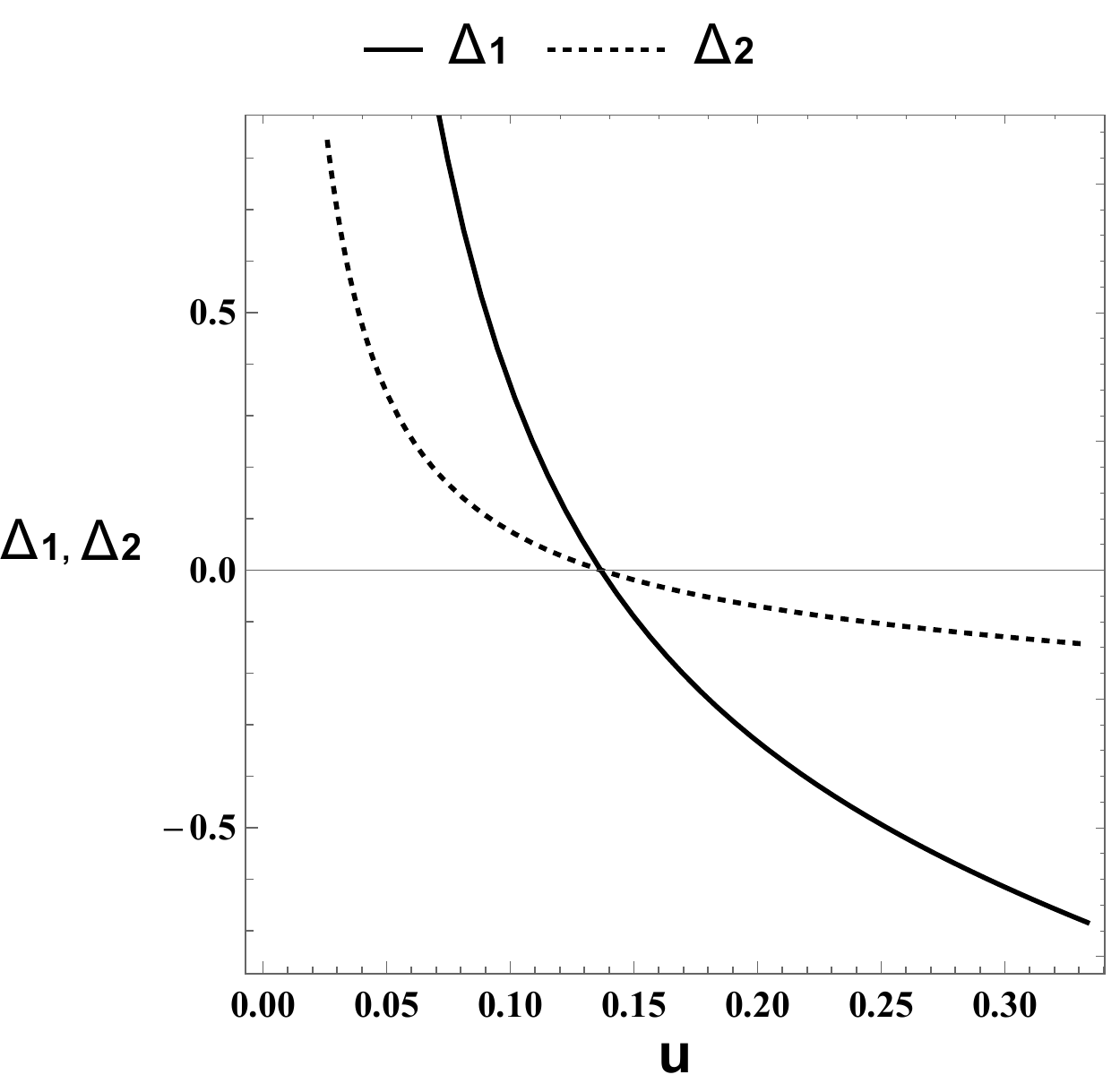}
\vspace{0mm}
\caption{$\Delta_1$ (solid line) and $\Delta_2$ (dotted line) as a function of the parameter $u$. 
$\vec{x}$ and $\vec{y}$ are the same as those used for Fig.~\ref{fig5}. 
In the range where $\Delta_1>0$, therefore $\Delta_2>0$, the non-trivial trade-off relation exists.}
\label{fig6}
\end{center}
\end{figure}

We can see that $\Delta_1$ and $\Delta_2$ in Fig.~\ref{fig5} indicates the ``strength" of trade-off relation 
by their definitions. They are calculated as 
\begin{align}
\Delta_1
&=\frac{\Delta}{g_\mathrm{S}^{\:22}-g_\mathrm{R}^{\:22}}=\frac{3}{4 \, \zeta |\vec{\eta}|^2 u (u^2-7 u +4)(3 u^2-9u+8)} F_\zeta(u), \nonumber \\
\Delta_2
&=\frac{\Delta}{g_\mathrm{S}^{\:11}-g_\mathrm{R}^{\:11}}=\frac{3}{4 \, \zeta |\vec{\xi}|^2 u (u^2-7 u +4)(3 u^2-9u+8)} F_\zeta(u). \nonumber 
\end{align}
The strengths of trade-off relation is proportional to $\Delta$. 
Figure~\ref{fig6} shows $\Delta_1$ and $\Delta_2$ as a function of the parameter $u$.  
In the range where $\Delta_1>0$ or $\Delta_2>0$, the non-trivial trade-off relation exists. 
The strength of trade-off relation becomes stronger as $u$ approaches 0. 
%
\subsection{Discussion}\label{sec:3} 
Unlike a qubit reference state or a pure state reference state, there exists a non-trivial trade-off relation 
for some qutrit reference states even when the generators commute.  
We show analytically that a non-trivial trade-off relation always exists in a certain range of the reference state parameter $u$ 
when the reference state $\rho_0$ is defined by Eq.~\eqref{single_u} that is 
a sum of the completely mixed state and a perturbation.
 
Furthermore, the strengths of trade-off relation $\Delta_1$ and $\Delta_2$ increase as $u$ approaches 0. This looks counterintuitive, 
because we can regard $u$ as a small perturbation from 3x3 identity matrix when $u \ll 1$ by the definition of $\rho_0$, Eq.~\eqref{single_u}.  
This reflects the fact that $\partial_i \rho_\theta$ is not necessarily small when the perturbation itself is small.  
Since the $(i,j)$ component of the RLD Fisher information matrix is $J_{\mathrm{R}, \, ij}=\tr[\partial_j \rho_\theta L_{R, \, i}^\dagger]$, 
the component $g_{\mathrm{R}, \, ij}$ may not be small 
if  $\partial_i \rho_\theta$ is not small. 

In a more general case when $\rho_0$ is expressed by Eq.~\eqref{ref_state}, we conducted numerical analysis. In this case also, 
there exists a non-trivial trade-off relation.  
Furthermore, in the case of four dimensional system with pure imaginary off-diagonal components, 
we also see a non-trivial trade-off relation by the same numerical analysis as well. 
With these, we conclude that the error trade-off relation is a generic phenomenon in the sense that it occurs with a finite volume in the spate space. 
\section{Conclusion}\label{sec:4} 
We have investigated whether the error trade-off relation exists in the generic two-parameter unitary models 
for finite dimensional systems with the commuting generators. 
By analyzing the necessary and sufficient conditions for the SLD and RLD CR bounds to intersect each other, 
we obtain the necessary and sufficient conditions for the existence of a non-trivial trade-off relation 
based on the SLD and RLD CR bounds for arbitrary finite dimensional system.

By using the conditions,  
we show two examples of the qutrit system with the non-trivial trade-off relation. 
The result of the reference state with multi-parameter indicates that the eigenvalues of the reference state be in a 
certain range. 
In the other model reference state with one-parameter, 
we show analytically that a non-trivial trade-off relation always exists in a certain range of the reference state parameter  
and that the region with the trade-off relation is up to about a half of the allowed region.

In our previous study about the trade-off relation of an infinite dimensional system~\cite{sf}, the bound is also given by 
both of the SLD and RLD CR bounds when the generators of the unitary transformation with the commuting generators.  
As shown in Figs.~\ref{fig5} and~\ref{fig6}, we confirmed that what we saw in our previous study is not special, but generic. 
When the reference state is a pure state or a general qubit state, 
we disprove the existence of a non-trivial trade-off relation.
\section*{Acknowledgment}
The work is partly supported by the FY2020 UEC Research Support Program, the University of Electro-Communications.  
\appendix
\section{Solution $u_0$ of $F_\zeta(u_0)=0$} \label{sec:sol_gts}
In this section, we investigate the solution $s_0$ of $F_\zeta(u_0)=0$. 
We check up to the fourth partial derivative of $F_\zeta(u)$ with respect to $s$ to see $F_\zeta(u)$ in the allowed range for $t$ and $s$.

Let $F_\zeta^{(n)}(u)=\dfrac{d^n F_\zeta(u)}{\partial u^n}$. 
Up to the fourth partial derivative of $F_\zeta(u)$ with respect to $u$ are as follows
\begin{align}
F_\zeta^{(1)}(u) 
&=-45 u^4 - 64 (1 + 7 \zeta) + 72 u^3 (3 + 8 \zeta)  \nonumber \\
&+ 32 u (9 + 61 \zeta) - 9 u^2 (43 + 224 \zeta),  \nonumber \\
F_\zeta^{(2)}(u)
&= 2 [ -90 u^3 + 108 u^2 (3 + 8 \zeta) + 16 (9 + 61 \zeta)  \nonumber \\
&- 9 u (43 + 224 \zeta)],  \nonumber \\
F_\zeta^{(3)}(u)
&= -18 [43 + 30 u^2 + 224 \zeta - 24 u (3 + 8 \zeta)],  \nonumber \\
F_\zeta^{(4)}(u)
&=-216 (-6 + 5 u - 16 \zeta).  \nonumber
\end{align}
$F_\zeta^{(3)}(u)$ is convex upward, because the coefficient of $u^2$ in $F_\zeta^{(3)}(u)$ is negative. 
Therefore, the extremum, in this case, the maximum of $F_\zeta^{(3)}(u)$ is given by
$u^{(4)}_0$ which is the solution of $F_\zeta^{(4)}(u^{(4)}_0)=0$.  The solution $u^{(4)}_0$ is given by 
\be
u^{(4)}_0=\frac{2}{5} (8 \zeta+3).  \nonumber
\ee
$u^{(4)}_0$ which gives the maximum of $F_\zeta^{(3)}(u)$ becomes minimum at $\zeta=0$. 
At $\zeta=0$, $u^{(4)}_0=6/5 = 1.2> 1/3$. 
Because of $u^{(4)}_0 > 1/3$, $F_\zeta^{(3)}(u)$ increases monotonically in the range $0<u<1/3$.

$F_\zeta^{(3)}(u)$ at $u=1/3$ is
\begin{align}
F_\zeta^{(3)}(\frac{1}{3})&=-6 (67 + 480 \zeta) < 0 \quad \mathrm{when} \quad (0< \zeta < \frac{1}{3}). \nonumber 
\end{align}
Then, we see $F_\zeta^{(3)}(u) < 0$ when $0<u<1/3$. 
Therefore, $F_\zeta^{(2)}(u)$ decreases monotonically when $0<u<1/3$. 

$F_\zeta^{(2)}(u)$ at $u=1/3$ is
\begin{align}
F_\zeta^{(2)}(\frac{1}{3}) &= \frac{286}{3} + 800 \zeta > 0 \quad \mathrm{when} \quad (0< \zeta < \frac{1}{3}). \nonumber
\end{align}
Therefore, $F_\zeta^{(1)}(u)$ increases monotonically when $0<u<1/3$. 
\begin{align}
F_\zeta^{(1)}(\frac{1}{3}) &=-\frac{32}{9} < 0 \quad \mathrm{when} \quad (0< \zeta < \frac{1}{3}). \nonumber
\end{align}
Therefore, $F_\zeta(u)$ decreases monotonically when $0<u<1/3$. 
The values of $F_\zeta(u)$ at the both ends, $u=0$ and $u=1/3$ are
\begin{align}
F_\zeta(0) &=64 \zeta , \nonumber \\
F_\zeta(\frac{1}{3}) &= -\frac{256}{27}.  \nonumber
\end{align}
With a given $\zeta$ in the range $0 < \zeta \leq 1/3$, there always exists only one solution $u_0$ that satisfies $F_\zeta(u_0)=0$ in 
the range $0 < u_0 \leq {1}/{3}$. 

\end{document}